\documentclass[11pt,twoside]{article}

\usepackage{asp2006}
\usepackage{epsf}
\usepackage{psfig}
\usepackage{lscape}

\begin{document}
\title{High precision millisecond pulsar timing with the EPTA}   %%% Fill in title
\author{K. Lazaridis}   %%% Fill in author names
\affil{Max-Planck-Institut f\"ur Radioastronomie, Auf dem H\"ugel 69,
  53121, Bonn, Germany}    %%% Fill in author affiliations

\begin{abstract} %%% Abstract to run on from here.
  The European Pulsar Timing Array (EPTA) network is a collaboration
  between the five largest radio telescopes in Europe aiming to study
  the astrophysics of millisecond pulsars and to detect cosmological
  gravitational waves in the nano-Hertz regime. The advantages and
  techniques of handling the multi-telescope datasets of a number of
  sources will be presented. In addition, the results of the EPTA
  timing analysis of the pulsar-white dwarf binary PSR J1012+5307 will
  be reported. Specifically, the measurements for the first time for
  this system, of the parallax, the variation of the projected
  semi-major axis and of the orbital period. Finally, the derived
  stringent, theory independent limits on alternative theories of
  gravity, with the use of this ideal laboratory for strong- field
  gravity tests, will be presented.
\end{abstract}

%%% MAIN BODY OF TEXT GOES HERE. CONSULT "INSTRUCTIONS FOR AUTHORS USING
%%% LATEX2E MARKUP", SECTIONS 2.3-2.6 FOR HELP WITH EQUATIONS, FIGURES,
%%% AND TABLES.

%\section{}   %%% Top level section head (remove "%" symbol)
%\subsection{}   %%% Second level section head (remove "%" symbol)
%\subsubsection{}   %%% Lowest level section head (remove "%" symbol)
%\section*{}    %%% Unnumbered top level section head (remove "%" symbol)
%\subsection*{}   %%% Unnumbered second level section head (remove "%" symbol)
\section{Introduction}
\label{sec:intro}

Millisecond pulsars (MSPs) are extremely stable and accurate cosmic
clocks due to their short spin periods and the high degree of
rotational stability. The technique of pulsar timing is the regular
monitoring of the rotation of these objects by tracking the times of
arrival (TOAs) of the radio pulses. Frequent measurements of pulse
arrival times of pulsars provide a powerful tool to determine their
spin and astrometric parameters to a very high degree of precision.

High precision timing of radio millisecond pulsars, most of which are
members of binary systems, can be the key to a number of unanswered
problems in fundamental physics and astronomy, ranging from stellar
evolution to tests of gravitational physics in the strong field regime
and the detection of a cosmological gravitational wave background.

\section{The European Pulsar Timing Array}
\label{sec:EPTA}

The European Pulsar Timing Array (EPTA) network (Table \ref{tab:EPTA})
is a collaboration between the five institutes (ASTRON, JBO, INAF,
MPIfR and Nan\c cay observatory) operating the largest radio
telescopes in Europe.  It is consisting of the Effelsberg 100m
radiotelescope of the Max-Planck-Institute for Radioastronomy (MPIfR)
in Germany, the 76m Lovell radiotelescope of the Manchester
University, at Jodrell Bank, UK, the 94m equivalent Westerbork
Synthesis Radio Telescope (WSRT) of ASTRON in the Netherlands, the 94m
equivalent Nan\c cay decimetric radio telescope (NRT) in France and
soon the 64m Sardinia Radio Telescope (SRT) in Italy.

The EPTA is using the available telescopes for high-precision timing
in a coordinated way, in schedules and source lists. This results, in
larger and denser datasets in multiple frequencies. In addition,
exchange of data, people and knowledge between the working groups is
ordinarily happening, resolving swiftly any systematic telescope
problems or other issues. The main aim of the EPTA is to increase the
precision and quality of pulsar timing measurements, to study the
astrophysics of millisecond pulsars and to detect cosmological
gravitational waves from coalescent massive binary black holes in the
nano-Hertz regime.

% ------------------------------------------------------------------------------------------------------------------------------------------------------------------------
\begin{table}[!ht]
\center
\caption{The European Pulsar Timing Array collaboration. On the first column the institutes and telescopes and on the second the people participating in the EPTA.}
\begin{tabular}{ll}
\hline
 {Institute--Telescope} & {People} \\
\noalign{\smallskip}
\hline
\noalign{\smallskip}
MPIfR--Effelsberg      &      M. Kramer, A. Jessner, N. Wex, \\
      &                              P. Freire, D. Champion, K. Lazaridis\\
  U. Manchester/Jodrell Bank    &    B. Stappers, A. Lyne,  C. Jordan, \\ 
    --Lovell &                                                   G. Janssen, M. Purver, S. Sanidas \\
  Nan\c cay observatory--NRT      &         I. Cognard, G. Theureau,      \\   
    &                                                    R. Ferdman, G. Desvignes\\ 
INAF--SRT     &          A. Possenti, M. Burgay,  M. Pilia  \\ 
ASTRON/U. Amsterdam/    &          J. Hessels, Y. Levin,     \\     
U. Leiden--Westerbork    &          R. van Haasteren           \\       
\noalign{\smallskip}
\hline
\end{tabular}
\label{tab:EPTA}
\end{table}
%---------------------------------------------------------------------------------------------------------------------------------------------------

Currently the EPTA is collaborating with the Parkes Pulsar Timing
Array (PPTA) in Australia and the nanoGrav in the USA, forming the
International Pulsar Timing Array (IPTA), a global pulsar timing array
with full sky coverage wishing to lead the way in the detection and
study of gravitational waves.

\subsection{Effelsberg millisecond pulsars}
\label{subsec:Eff}

In the current work, several millisecond pulsars have been chosen from
the Effelsberg source list.  On the archival data of those, new
calibration techniques were applied \citep{laz09}, i.e. ephemeris
update, noise removal from individual channels, new time corrections,
summation of individual scans and creation of new synthetic templates
\citep{kwj+94}. This re-calibration resulted in an improved timing
accuracy making the pulsars ideal candidates in the effort of directly
detecting gravitational waves in the nano-Hertz regime. In total
fifteen sources, observed monthly with the Effelsberg radio telescope
at at least two different frequencies, have been chosen. In Table
\ref{tab:14pulsars} the selected sources are shown with the current
post-fit rms of the model fit achieved, in comparison with the
post-fit rms without the improved calibration procedures. In all the
cases the improvement is vast, varying from two times to two orders of
magnitude. In the last column the post-fit rms with the use of only
the 1.4GHz data is shown.
% ------------------------------------------------------------------------------------------------------------------------------------------------------------------------
\begin{table}[!ht]
\center
\caption{The fifteen analyzed millisecond pulsars with the post-fit rms with and without calibration. The TOAs are until September 2008 (MJD 54720) and all the available Effelsberg frequencies have been used. On the last column only the 1.4\,GHz TOAs have been used.}
\begin{tabular}{cccc}
\hline
 {Source} &{Post-fit rms} & {Post-fit rms}   & {Post-fit rms} \\
              &   {before ($\mu$s)}       &    {after ($\mu$s)}  & {1.4\,GHz ($\mu$s)} \\
\noalign{\smallskip}
\hline
\noalign{\smallskip}
PSR J0030+0451     &   ----- $^a$      &      5.7       &  3.8 \\   
PSR J0218+4232     &    51.5     &          9.6          &  9.1   \\ 
PSR J0613$-$0200     &    28.1      &         2.7        &   2.6   \\   
PSR J0621+1002    &     23.8     &          6.8          &   6.5\\     
PSR J0751+1807    &     15.5     &          4.9        &   4.3\\
PSR J1012+5307   &      3.5       &         2.7        &   2.6\\
PSR J1022+1001    &     16.2     &          3.7        & 3.1\\          
PSR J1024$-$0719     &    $\sim$1500       &        13.4       &  2.7\\
PSR J1518+4904    &     23.5     &         18.9         &   19.0  \\ 
PSR J1623$-$2631     &    $\sim$2100       &         3.9        &  3.9 \\    
PSR J1640+2224    &     15.2    &          1.7        & 1.5\\   
PSR J1643$-$1224     &    34.4      &         3.9       & 3.8    \\     
PSR J1744$-$1134     &     1.7       &        0.6         &  0.6        \\
PSR J2051$-$0827     &    49.9       &       48.7      &   48.8    \\
PSR J2145$-$0750     &     4.1       &        2.6      &  2.5\\
\noalign{\smallskip}
\hline
\multicolumn{4}{l}{\footnotesize{$^a$: The source was regularly observed from 2008.}}
\end{tabular}
\label{tab:14pulsars}
\end{table}
%---------------------------------------------------------------------------------------------------------------------------------------------------

At least twelve of the analyzed ms pulsars can be good candidates for
the EPTA efforts in detecting the stochastic background of
gravitational radiation. Work is in progress for combination of the
datasets from all the EPTA telescopes for all of these sources, which
will improve the current measurements and take us closer to the
desired detection limit. Until reaching this point the combined
datasets are being used for doing high precision timing analysis of
the individual systems, as it is shown in the following section for
PSR J1012+5307.

\section{PSR J1012+5307}
\label{sec:1012}

\subsection{Introduction}
\label{subsec:intro1012}

PSR J1012+5307 is a 5.3\,ms pulsar in a low eccentricity binary system
with orbital period of $P_b=$14.5\,h \citep{nll+95} and a low mass
helium white dwarf (WD) companion \citep{llfn95}. \cite{cgk98} compared the
measured optical luminosity of the WD to the value expected from WD
models and calculated a distance of $d = 840 \pm 90$\,pc. In addition
they measured, a radial velocity component of $44 \pm 8$\,km\,s$^{-1}$
relative to the solar system barycenter (SSB), and the mass ratio of
the pulsar and its companion $q=m_{p}/m_{c} = 10.5\pm 0.5$. Finally
they derived a companion mass of $m_{c} = 0.16 \pm 0.02\,M_{\odot}$, a
pulsar mass of $m_{p} = 1.64\pm 0.22\,M_{\odot}$ and an orbital
inclination angle of $i=52^\circ \pm 4^\circ$.

\cite{lcw+01} presented a complete precision timing analysis of PSR
J1012+5307 using 4 years of timing data from Effelsberg and 7 years
from Lovell radio telescope. They derived the spin, astrometric and
binary parameters for the system and they discussed the prospects of
future measurements of Post-Keplerian parameters (PK) which can
contribute to the derivation of stringent limits on alternative
gravity theories.

In this work PSR J1012+5307 has been revisited with seven
more years of high-precision timing data and combined
datasets from the EPTA
telescopes. The data have been analyzed using the 
timing software TEMPO\footnote{http://www.atnf.csiro.au/research/pulsar/tempo/}. 

\subsection{Results}
\label{subsec:results1012}

\begin{figure}[!ht]
\plotfiddle{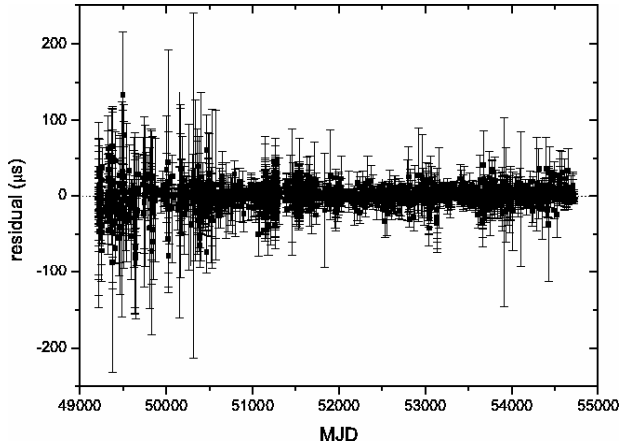}{2.2in}{0}{35}{35}{-120}{-130}
\caption{Post-fit timing residuals using datasets from Effelsberg, Lovell, Nan\c cay and Westerbork.}
\label{fig:postfit}
\end{figure}

Combining almost 3000 TOAs, at five different frequencies from the four
telescopes now in use by the EPTA, we have been able to improve on the 
timing solution and on all the astrometric, spin and binary 
parameters of this system, see Figure \ref{fig:postfit}.

%-------------------------------------------------------------------------------
%-------------------------------------------------------------------------------

For the first time a parallax $\pi = 1.2\pm 0.3$\,mas has been
measured for PSR J1012+5307. This corresponds to a distance of $d =
822 \pm 178$\,pc which is consistent with the $d = 840 \pm 90$ pc
measured from the optical observations. In addition, the proper motion
measurements have been improved by an order of magnitude, with a total
proper motion of $\mu_{t}=25.735\pm0.019$\,mas\,yr$^{-1}$ yielding a
total transverse velocity of $v_t = 102.0\pm 9.8$\,km\,s$^{-1}$. Using
the radial velocity from the optical measurements, the space velocity
of the system $v_{space} = 111.4 \pm 9.5$\,km\,s$^{-1}$ has been
calculated, being consistent and almost three times more precise than
the previous value.

An upper limit of $e < 8.4\times 10^{-7} (95\;{\rm per\;cent}\;{\rm
  C.L.})$ has been obtained, being consistent with the evolutionary
scenario of spin-up of PSR J1012+5307 through mass transfer from the
companion, while it is in the red giant phase. A change in the
projected semi-major axis of $\dot{x}_{obs} = 2.3(8)\times 10^{-15}$
has been measured in the current analysis, which has been used to
constrain for the first time the position angle of the ascending node.

As predicted by \cite{lcw+01} a significant measurement of the change
in the orbital period of the system, $\dot{P_b}=5.0(1.4) \times
10^{-14}$, has been obtained for the first time. This is caused by the
Doppler correction, which is the combined effect of the proper motion
of the system \citep{shk70} and a correction term for the Galactic
acceleration and by a contribution due to the quadrupole term of the
gravitational wave emission, as predicted by general relativity
(GR). After subtracting these two contributions from our measured
value, the excess value of $\dot{P}_b^{exc} = (-0.4 \pm 1.6) \times
10^{-14}$ confirms the validity of GR for one more millisecond pulsar
binary system.

All the terms mentioned above are the ones expected to contribute by
using GR as our theory of gravity. However, there are alternative
theories of gravity, that violate the strong equivalence principle (SEP) and
predict extra contributions to the observed orbital period
variation. One is the dipole term of the gravitational wave emission,
which results from the difference in gravitational binding energy of
the two bodies of a binary system. Thus, the case of PSR J1012+5307,
where there is a pulsar-WD system, is ideal for testing the strength
of such emission. One finds for small-eccentricity pulsar-WD systems,
where the sensitivity $s$ (related to the gravitational self-energy of
a body) of the WD is much smaller than the one of the pulsar,
$\dot{P_{b}}^{dipole} = -4\pi^2 \, \frac{T_\odot \mu}{P_b} \,
\kappa_{D} {s_p}^{2}$, \citep{wil01} where $T_\odot = 4.9255\,\mu s$
and $\mu$ is the reduced mass. $s_p$ is the sensitivity of the pulsar
and $\kappa_{D}$ refers to the dipole self-gravitational contribution,
which takes different values for different theories of gravity (zero
for GR).

Another term is predicted by the variation of the locally measured
gravitational constant as the universe expands, $\dot{P}_b^{\dot{G}} =
-2\,\frac{\dot{G}}{G} \left[1 - \left(1+\frac{m_c}{2M}\right)
  s_p\right] P_{b} $, \citep{dgt88, nor90} where $M$ is the total mass
of the system. It has been shown that there is no need to add these
extra contributions to explain the variations of the orbital period,
however the excess value has been used to set limits for a wide class
of alternative theories of gravity.

PSR J1012+5307 is an ideal lab for constraining the dipole radiation
term because the WD nature of the companion is affirmed optically, the
mass estimates are free of any explicit strong-field effects and the
mass of the pulsar is rather high, which is important in the case of
strong field effects that occur only above a certain critical mass,
like the spontaneous scalarisation \citep{de93}. Thus, by using the
$\dot{G}/G = (4 \pm 9) \times 10^{-13}$\,yr$^{-1}$ limit from the Lunar
Laser Ranging (LLR) \citep{wtb04} the $\dot{G}/G$ contribution has been
calculated and subtracted from our excess value in order to finally
obtain an improved generic limit for the dipole contribution of
$\kappa_D = (0.2 \pm 2.4) \times 10^{-3}$ (95 per cent C.L.).

\begin{figure}[!ht]
\plotfiddle{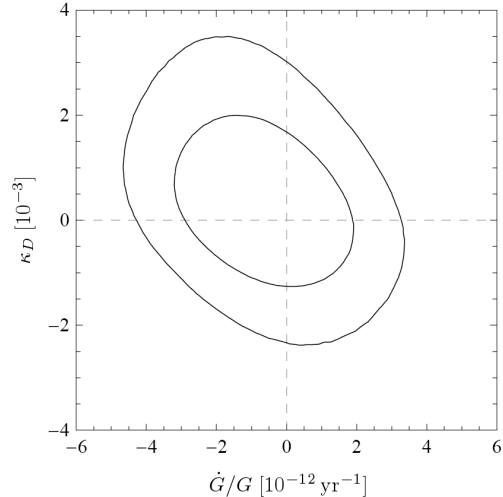}{2.2in}{0}{40}{40}{-110}{-20}
\caption{Contour plots of the one and two $\sigma$ confidence regions on $\dot
  G/G$ and $\kappa_D$ jointly. The elongation of the regions reflects the
  correlation due to the mutual dependence of the two pulsar-WD systems
 in this combined test.}
\label{fig:monteKG}
\end{figure}

A generic test for $\dot G$ cannot be done with a single binary
pulsar, since general theories that predict a variation of the
gravitational constant typically also predict the existence of dipole
radiation. This degeneracy has been broken here in a joint analysis of
PSR J1012+5307 and PSR J0437$-$4715 \citep{vbv+08}, two binary
pulsar-WD systems with tight limits for ${\dot P}_b$ and different
orbital periods. By applying equation $ \frac{\dot{P}_b^{exc}}{P_b} =
- 2 \frac{\dot G}{G} \left[1 - \left(1+\frac{m_c}{2M}\right)
  s_p\right] - 4 \pi^2 \frac{T_\odot\mu}{P_b^2} \, \kappa_{\rm D}
s_p^2 \nonumber\\$ to both binary pulsars, and solving in a
Monte-Carlo simulation (Figure \ref{fig:monteKG}) this set of two
equations, stringent and generic limits based purely on pulsar data
and in the strong field regime have been obtained. With a 95 per cent
C.L., $\frac{\dot G}{G} = (-0.7 \pm 3.3) \times 10^{-12} \; {\rm
  yr}^{-1}$ and $\kappa_{\rm D} = (0.3 \pm 2.5) \times 10^{-3}$.  In
the future, more accurate measurements of $\dot{P_b}$ and distance of
the two pulsars and WDs could constrain even more our derived limits.

%-------------------------------------------------------------------------------

A more detailed work on this study of
PSR J1012+5307 can be found in \cite{lwj+09}.

\acknowledgements %%% Text of acknowledgements runs on after this command.
We are grateful to all the EPTA collaborators for providing data, for
valuable discussions and for their overall help.  ln addition, to all
staff at the EPTA radio telescopes for their help with the
observations. Kosmas Lazaridis was supported for this research through
a stipend from the International Max Planck Research School (IMPRS)
for Astronomy and Astrophysics at the Universities of Bonn and
Cologne.

%%% THE BIBLIOGRAPHY
%%%
%%% CONSULT SECTION 3 OF "INSTRUCTIONS FOR AUTHORS" FOR HOW TO USE NATBIB.
%%% AUTHORS ARE ENCOURAGED TO USE EITHER THE "THEBIBLIOGRAPY" ENVIRONMENT
%%% BY UNCOMMENTING (DELETING THE "%" SYMBOL) THE COMMANDS BELOW, OR BY
%%% USING THE BIBTEX ENVIRONMENT. TO FIND OUT WHICH IS APPLICABLE TO YOUR
%%% CONTRIBUTION, CONSULT THE VOLUME EDITORS FOR YOUR PROCEEDINGS.
%%%


\begin{thebibliography}{}
\bibitem[{Callanan {et~al.}(1998)Callanan, Garnavich, \& Koester}]{cgk98}
Callanan, P.~J., Garnavich, P.~M., \& Koester, D. 1998, \mnras, 298, 207
\bibitem[{Damour \& Esposito-Farese(1993)}]{de93}
Damour, T. \& Esposito-Farese, G. 1993, \prl, 70, 2220
\bibitem[{Damour {et~al.}(1988)Damour, Gibbons, \& Taylor}]{dgt88}
Damour, T., Gibbons, G.~W., \& Taylor, J.~H. 1988, \prl, 61, 1151
\bibitem[{Kramer {et~al.}(1994)Kramer, Wielebinski, Jessner, Gil, \&
  Seiradakis}]{kwj+94}
Kramer, M., Wielebinski, R., Jessner, A., Gil, J.~A., \& Seiradakis, J.~H.
  1994, \aaps, 107, 515
\bibitem[{Lange {et~al.}(2001)Lange, Camilo, Wex, Kramer, Backer, Lyne, \&
  Doroshenko}]{lcw+01}
Lange, C., Camilo, F., Wex, N., {et~al.} 2001, \mnras, 326, 274
\bibitem[{{Lazaridis}(2009)}]{laz09}
{Lazaridis}, K. 2009, Ph.D.~Thesis
\bibitem[{Lazaridis {et~al.}(2009)Lazaridis, Wex, Jessner, Kramer,
    Stappers, Janssen, Desvignes, Purver, Cognard, Theureau, Lyne,
    Jordan, Zensus}]{lwj+09} {Lazaridis}, K., {Wex}, N., {Jessner},
  A., {Kramer}, M., {Stappers}, B.~W., {Janssen}, G.~H., {Desvignes},
  G., {Purver}, M.~B., {Cognard}, I., {Theureau}, G., {Lyne}, A.~G.,
  {Jordan}, C.~A. and {Zensus}, J.~A. 2009, \mnras, 400, 805
\bibitem[{Lorimer {et~al.}(1995{\natexlab{a}})Lorimer, Lyne, Festin, \&
  Nicastro}]{llfn95}
Lorimer, D.~R., Lyne, A.~G., Festin, L., \& Nicastro, L. 1995{\natexlab{a}},
  \nat, 376, 393
\bibitem[{Nicastro {et~al.}(1995)Nicastro, Lyne, Lorimer, Harrison, Bailes, \&
  Skidmore}]{nll+95}
Nicastro, L., Lyne, A.~G., Lorimer, D.~R., {et~al.} 1995, \mnras, 273, L68
\bibitem[{Nordtvedt(1990)}]{nor90}
Nordtvedt, K. 1990, \prl, 65, 953
\bibitem[{Shklovskii(1970)}]{shk70}
Shklovskii, I.~S. 1970, \sovast, 13, 562
\bibitem[{{Verbiest} {et~al.}(2008){Verbiest}, {Bailes}, {van Straten}, \&
  et~al.}]{vbv+08}
{Verbiest}, J.~P.~W., {Bailes}, M., {van Straten}, W., \& et~al. 2008, \apj,
  679, 675
\bibitem[{{Will}(2001)}]{wil01}
{Will}, C. 2001, Living Reviews in Relativity, 4, 1, uRL (Cited on 2006/02/01):
  http://www.livingreviews.org/Irr-2001-4
\bibitem[{Williams {et~al.}(2004)Williams, Turyshev, \& Boggs}]{wtb04}
Williams, J.~G., Turyshev, S.~G., \& Boggs, D.~H. 2004, \prl, 93, 261101
\end{thebibliography}
\end{document}